\documentclass[reprint, twocolumn, amssymb, amsmath, bibnotes, aps, prl, superscriptaddress,longbibliography]{revtex4-2} 

\usepackage{graphicx,comment}
\usepackage{float}
\usepackage{dcolumn}
\usepackage{bm}
\usepackage[dvipsnames]{xcolor}
\usepackage[colorlinks=true]{hyperref}
\definecolor{darkblue}{RGB}{33,33,137}
\definecolor{darkred}{RGB}{193,23,23}
\hypersetup{
    colorlinks=true,       
    linkcolor=blue,     
    citecolor=blue,    
    urlcolor=blue      
} 
\usepackage{natbib}
\usepackage[caption=false]{subfig}
\usepackage{color,soul}
\usepackage{tikz,fp}
\usepackage{tikz-cd}
\usepackage{environ}
\usepackage{upgreek} 
\usetikzlibrary{arrows}
\usetikzlibrary{intersections}
\usetikzlibrary{shapes.geometric}
\usetikzlibrary{decorations.pathmorphing, patterns,shapes,fixedpointarithmetic}
\usetikzlibrary{decorations.markings}

\pgfdeclarepatternformonly{south west lines}{\pgfqpoint{-0pt}{-0pt}}{\pgfqpoint{3pt}{3pt}}{\pgfqpoint{3pt}{3pt}}{
  \pgfsetlinewidth{0.4pt}
  \pgfpathmoveto{\pgfqpoint{0pt}{0pt}}
  \pgfpathlineto{\pgfqpoint{3pt}{3pt}}
  \pgfpathmoveto{\pgfqpoint{2.8pt}{-.2pt}}
  \pgfpathlineto{\pgfqpoint{3.2pt}{.2pt}}
  \pgfpathmoveto{\pgfqpoint{-.2pt}{2.8pt}}
  \pgfpathlineto{\pgfqpoint{.2pt}{3.2pt}}
  \pgfusepath{stroke}}

\NewEnviron{myequation}{%
    \begin{equation}
    \scalebox{0.93}{$\displaystyle{\BODY}$}
    \end{equation}
    }

\tikzset{
  mid arrow/.style={postaction={decorate,decoration={
        markings,
        mark=at position .575 with {\arrow{stealth}}
  }}},
  near arrow/.style={postaction={decorate,decoration={
        markings,
        mark=at position .275 with {\arrow{stealth}}
  }}},
  far arrow/.style={postaction={decorate,decoration={
        markings,
        mark=at position .800 with {\arrow{stealth}}
  }}},
  snake arrow/.style={fixed point arithmetic, decorate, decoration={snake,amplitude=2pt, segment length=11pt},postaction={decoration={markings,mark=at position 0.625 with {\arrow{stealth}}},decorate}},
}
\tikzset{
  baseline = -0.5ex,
  wavy/.style = {
    thick,
    decorate,
    decoration={snake,amplitude=2pt,segment length=5pt}},
  sdot/.style = {
    circle,
    draw=none,
    fill=black,
    minimum size=2.5pt,
    inner sep=0pt},
  bdot/.style = {
    circle,
    draw=none,
    fill=black,
    minimum size=4pt,
    inner sep=0pt},
  svertex/.style = {
    circle,
    draw=black,
    thick,
    fill=lightgray,
    minimum size=8pt,
    inner sep=1pt},
  bvertex/.style = {
    circle,
    draw=black,
    thick,
    fill=lightgray,
    minimum size=24pt},
  bvertexsmall/.style = {
    circle,
    draw=black,
    thick,
    fill=lightgray,
    minimum size=7pt},
  bvertexnormal/.style = {
    circle,
    draw=black,
    thick,
    fill=lightgray,
    minimum size=16pt},
  dvertex/.style = {
    circle,
    draw=black,
    thick,
    fill=gray,
    minimum size=25pt}}

\usepackage{pdfpages} 
\usepackage{pgffor} 

\makeatletter
\AtBeginDocument{\let\LS@rot\@undefined}
\makeatother

\def\supplementfilename{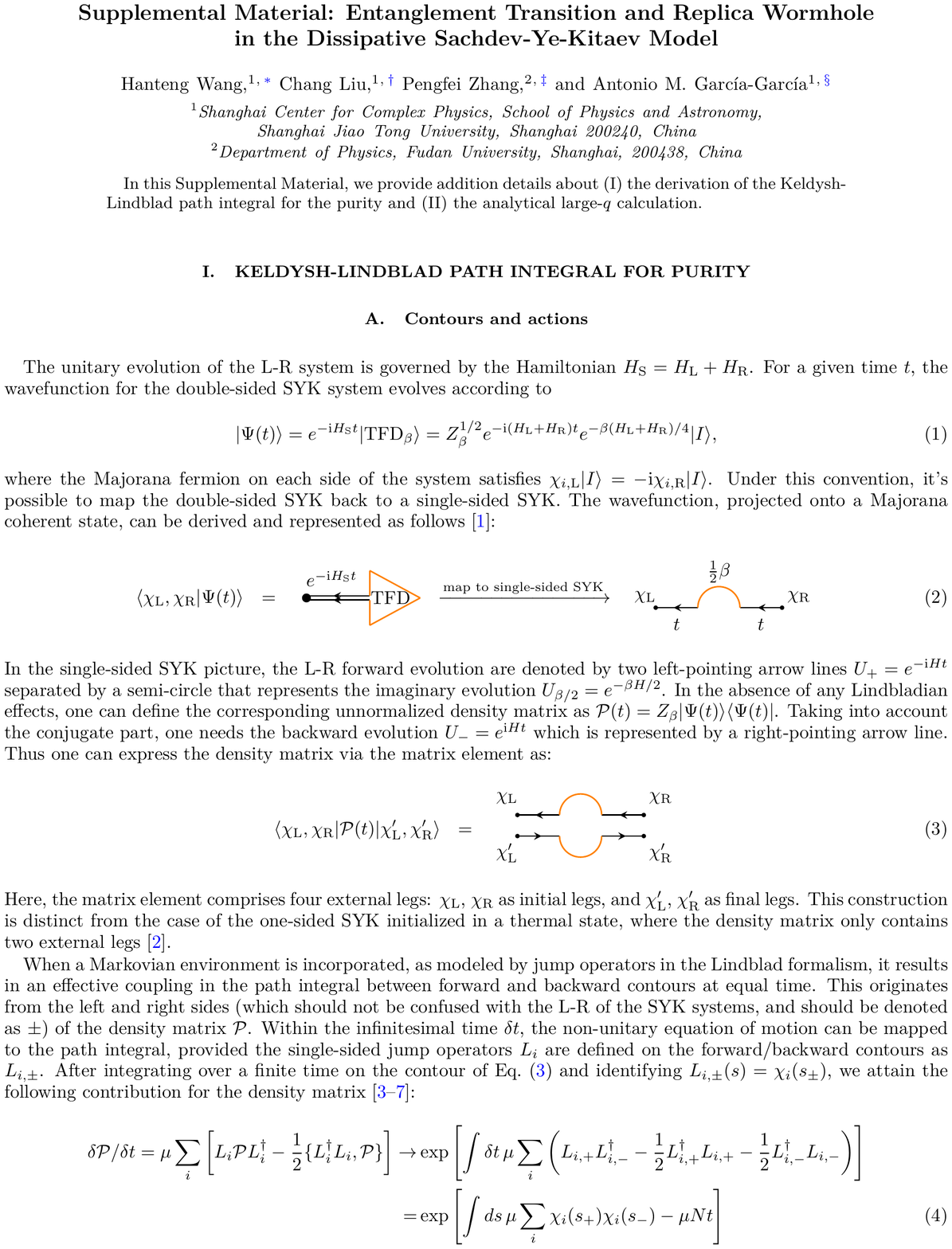}

\pdfximage{\supplementfilename}
\def\numbersupplementpages{\the\pdflastximagepages}

\newif\ifarXiv
\arXivtrue 

\begin{document}

\title{Entanglement Transition and Replica Wormhole in the Dissipative \\ Sachdev-Ye-Kitaev Model}
\author{Hanteng Wang}
\thanks{wanghanteng@sjtu.edu.cn}
\affiliation{%
Shanghai Center for Complex Physics, School of Physics and Astronomy, Shanghai Jiao Tong University, Shanghai 200240, China
}%

\author{Chang Liu}
\thanks{cl91tp@gmail.com}
\affiliation{%
Shanghai Center for Complex Physics, School of Physics and Astronomy, Shanghai Jiao Tong University, Shanghai 200240, China
}%

\author{Pengfei Zhang}
\thanks{pengfeizhang.physics@gmail.com}
\affiliation{%
Department of Physics, Fudan University, Shanghai, 200438, China
}%

\author{Antonio M. García-García}
\thanks{amgg@sjtu.edu.cn}
\affiliation{%
Shanghai Center for Complex Physics, School of Physics and Astronomy, Shanghai Jiao Tong University, Shanghai 200240, China
}%

\date{\today}

\begin{abstract}
Recent discoveries have highlighted the significance of replica wormholes in resolving the information paradox and establishing the unitarity of black hole evaporation. In this letter, we propose the dissipative Sachdev-Ye-Kitaev model (SYK) as a minimal quantum model that exhibits entanglement dynamics with features qualitatively similar to replica wormholes. As a demonstration, we investigate the entanglement growth of a pair of dissipative SYK models initialized in a thermofield double state (TFD). In the regime of large $N$ with weak dissipation, we observe a first-order entanglement transition characterized by a switch of the dominant saddle point: from replica diagonal solutions for short times to replica wormhole-like off-diagonal solutions for long times. Furthermore, we show that signature of replica wormholes persists even at moderate $N \lesssim 30$ by using the Monte Carlo quantum trajectory method. Our work paves the way for explorations of replica wormhole physics in quantum simulators. 
\end{abstract}

\maketitle
Recently, there has been a growing interest in comprehending systems with non-unitary dynamics. Experimental investigations into $PT$-symmetry breaking of non-Hermitian Hamiltonians have been conducted in the field of quantum optics \cite{guo2010} and cold atoms \cite{li2019}. Additionally, numerous theoretical predictions have been made, encompassing studies on information flow \cite{kawabata2017}, non-Hermitian superconductivity \cite{kawabata2018}, and particle detectors \cite{wiersig2014}. Significant progress has also been achieved on the role of dissipation in quantum many-body dynamics \cite{zhai2020a, garcia2023a, li2021a, xu2020, cornelius2022, xu2019}, which unveils enriched symmetry classes \cite{ueda2019, garcia2022d, kawabata2022a, sa2022a}. Moreover, the non-unitarity can drive dynamical transitions when following quantum trajectories
\cite{Li:2018mcv,Skinner:2018tjl,Chan:2018upn,choi2020quantum} or maintaining access to different environments \cite{weinstein2022, zhang2022a}.

Interestingly, the study of non-unitary dynamics are gaining traction in a completely different field: quantum gravity. On the one hand, this benefits from the existence of a low-energy duality between the Sachdev-Ye-Kitaev (SYK) model \cite{sachdev1993,kitaev2015,maldacena2016,Kitaev:2017awl} and Jackiw-Teitelboim gravity \cite{jackiw1985,teitelboim1983,almheiri2015,maldacena2016a}, which enables the study of black holes and wormholes in concrete quantum systems \cite{Maldacena:2018lmt,PhysRevLett.124.221601,Saad:2018bqo,Qi:2020ian,garcia2021,garcia2022a,kulkarni2022,kawabata2022,garcia2022e,PhysRevResearch.4.L022068}. On the other hand, the coupling of black holes to an environment is naturally connected to the so called information paradox \cite{hawking1976}. The recent progress towards the resolution of the paradox identifies a new Ryu-Takayanagi (RT) surface \cite{Ryu:2006bv,Ryu:2006ef,Lewkowycz:2013nqa} with entanglement islands, which can be traced back to novel complex saddles known as replica wormholes \cite{almheiri2020,penington2020,almheiri2020a}. This leads to a first order Page transition \cite{page1993} in the growth of entanglement. Similar features have been observed in SYK settings where the environment is modeled as a Majorana chain \cite{chen2020replica}. 

\begin{figure}[t]
  \centering
  \includegraphics[width=0.95\columnwidth]{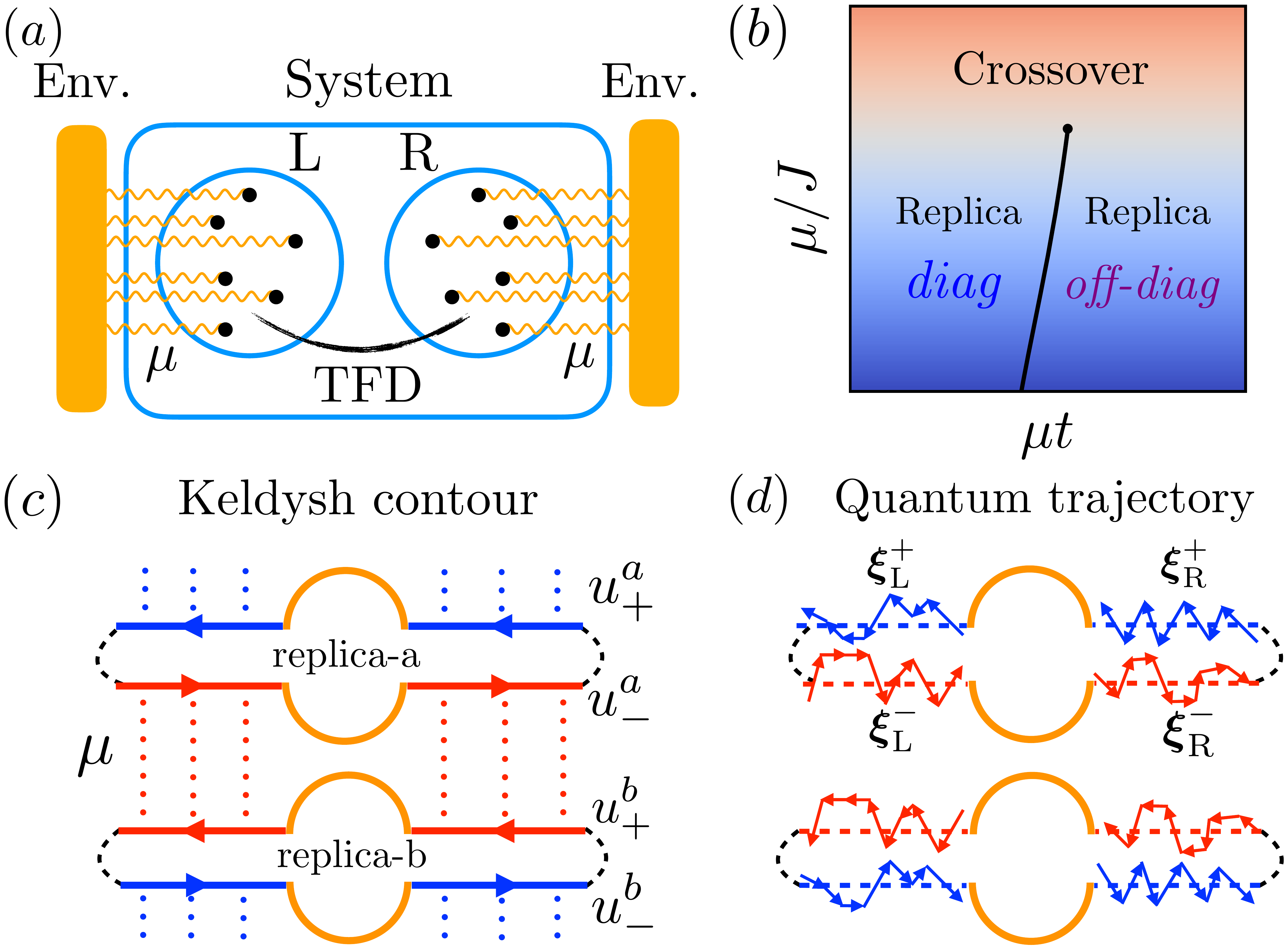}
  \caption{(a) Illustration of the setup. L and R systems are entangled without direct interaction between them. (b) Dynamical phase diagram of the system. The x-axis stands for real time in units of $1/\mu$, and the y-axis stands for the coupling strength $\mu$ in unit of $J$. (c) Sketch of the purity calculation in replica a,b with four time contours $u_+^a$, $u_-^a$, $u_+^b$, and $u_-^b$, connected by the Lindbladian coupling $\mu$. (d) An example of a quantum trajectory employed in the purity calculation at finite $N$.}
  \label{fig:setup}
\end{figure}

SYK-like models have been implemented in quantum simulation platforms \cite{Luo:2017bno,Jafferis:2022crx}. Consequently, SYK systems coupled to an environment becomes a natural candidate for observing replica wormholes. However, models incorporating microscopic baths \cite{Chen:2017dbb,Zhang:2019fcy,Almheiri:2019jqq,chen2020replica} are unfavorable for quantum simulations due to their requirement for additional logical qubits. 
In this letter, we investigate whether the signature of replica wormholes exists in a simpler setup where this problem does not arise: the dissipative SYK model \cite{kulkarni2022,kawabata2022,garcia2022e,PhysRevResearch.4.L022068}, where the bath's influence is modeled using Lindblad jump operators \cite{lindblad1976,sieberer2016keldysh}. We demonstrate that the entanglement transition can still be observed, provided that the dissipation strength is sufficiently weak, even in systems with a moderate number of fermions $N \lesssim 30$.  Figure~\ref{fig:setup} provides a schematic of the dynamical phase diagram. This finding paves the way for potential experimental exploration of quantum gravity via the utilization of strongly interacting dissipative quantum matter.

{\em\color{blue} Model.---}We consider a pair of $q$-body SYK models with identical internal couplings, denoted by L and R. Each of them is described by the Hamiltonian,
\begin{align}\label{Hamiltonian}
H= \!\!\!\! \sum_{1\leq i_1< i_2< ...< i_q\leq N} \!\!\!\! \mathrm{i}^{q/2} J_{i_1i_2...i_q}\chi_{i_1}\chi_{i_2}...\chi_{i_q},
\end{align}
where $N$ is the number of Majoranas defined by $\{\chi_i, \chi_j\}=\delta_{ij}$, $J_{i_1i_2...i_q}$ are Gaussian independent random couplings with zero mean and variance $\langle J_{i_1i_2...i_q}^2\rangle_{\text{dis}} = (q-1)!J^2/N^{q-1}$, and $\mathrm{i}$ denotes the imaginary unit. The single-sided SYK Hamiltonian $H$ has $2^{N/2}$ eigenstates $|n\rangle$ with energies $E_n$. The full two-sided Hamiltonian of the system can be represented as $H_\text{S}=H_\mathrm{L}+H_\mathrm{R}=H\otimes\mathbb{I}+\mathbb{I}\otimes H^T$. The L-R systems are initially highly entangled and prepared in a TFD state with inverse temperature $\beta$ in the form \cite{Israel:1976ur}:
\begin{align}
\label{eq:TFD}
|\Psi(t=0)\rangle = |\mathrm{TFD}_{\beta}\rangle = Z_\beta^{-1/2} e^{-\frac{\beta}{4}(H_\text{L}+H_\text{R})}|I\rangle,
\end{align}
where $|I\rangle$ is the maximally entangled state satisfying $\chi_{i,\mathrm{L}}|I\rangle = -\mathrm{i}\chi_{i,\mathrm{R}}|I\rangle$ \cite{kulkarni2022,gu2017}. The single SYK thermal partition function $Z_\beta = \mathrm{Tr}(e^{-\beta H})$ serves as the normalization.

The growth of entanglement has also been studied in SYK-like systems \cite{gu2017,pengfei2021b,zhang2020entanglement,Haldar:2020ymg,Chen:2020atj,Chen:2019qqe,dadras2021perturbative,jian2021phase,Dadras:2019tcz,Zhang:2023vpm}, including in studies of measurement induced phase transitions \cite{pengfei2021c,liu2021non,zhang2021emergent,milekhin2022,Tian-GangZhou:2023ghw}. However, most of these works focus on closed systems or microscopic environment models. In this paper, we instead consider a Markovian environment \cite{kawabata2022,garcia2022e,PhysRevResearch.4.L022068}, which results in the Lindbladian form of the density matrix $\mathcal{P}(t)$ time evolution:
\begin{equation}
\begin{aligned}
\label{eq:Lindblad}
\partial_t\mathcal{P}= &-\mathrm{i}[H_\text{S},\mathcal{P}]\\
&+\sum_{i}^{N} \sum_{x=\text{L},\text{R}} \left[L_{x,i}\mathcal{P} L_{x,i}^{\dagger} - \frac{1}{2}\left\{\mathcal{P}, L_{x,i}^\dagger L_{x,i}\right\}\right],
\end{aligned}
\end{equation}
with jump operators $L_{\text{L},i} = \sqrt{\mu}\,\chi_{\text{L},i}$ and $L_{\text{R},i} = \sqrt{\mu}\,\chi_{\text{R},i}$. The coupling strength $\mu$ between the systems and their environment is identical for all sites $i$.

The aim of this work is to study the dynamical entanglement structure of this dissipative SYK model with TFD initial states by calculating its second Rényi entropy, $S = -\langle \ln \gamma\rangle_\text{dis}$, where $\gamma(t) = \frac{\mathrm{Tr(\mathcal{P}^2)}}{(\mathrm{Tr\mathcal{P})}^2}$ is called purity, and $\langle\cdot\cdot\cdot\rangle_{\mathrm{dis}}$ denotes an average over different realizations of SYKs. For the sake of simplicity, we use an unnormalized density matrix $\mathrm{Tr\mathcal{P}} = Z_\beta$. Due to the self-averaging feature of the SYK model \cite{wang2019replica,Kitaev:2017awl,Gu:2019jub,baldwin2020quenched}, we compute annealed instead of quenched averages, i.e., $S \approx -\ln \langle \gamma\rangle_\text{dis}$. 

{\em \color{blue}Path Integral for Purity.---}
We use the Keldysh path integral formulation \cite{kamenev2011field,sieberer2016keldysh,thompson2023field} to evaluate the dynamics of the unnormalized purity $\mathrm{Tr}(\mathcal{P}^2)$. We introduce two replicas, $a$ and $b$. As a result, four distinct time contours are needed, defined as $u_{+}^a$, $u_{-}^a$, $u_{+}^b$, and $u_{-}^b$, which can be compactified in terms of a single time variable $u$ as illustrated in Fig.~\ref{fig:setup}(c). As a result, $\mathrm{Tr}(\mathcal{P}^2)$ reads \cite{SM}:
\begin{myequation}
\begin{aligned}
 \mathrm{Tr}(\mathcal{P}^2) = \int&\mathcal{D}\chi\exp\Bigg\{-\int_{\mathcal{C}} du \left[\frac{1}{2}\sum_i^N\chi_i\partial_u\chi_i +f(u)H\right] \\
 &+\mu\int du \sum_i^N \sum_{\eta\neq\eta^\prime}\chi_i(u_\eta^a)\chi_i(u_{\eta^\prime}^b) - 2\mu Nt\Bigg\}.
\end{aligned}
\end{myequation}
The first line in the exponential represents the unitary evolution where $f(u)=\mathrm{i},\,-\mathrm{i},\,1$ corresponds to forward ($+$), backward ($-$), and rotational (imaginary) evolution contours respectively. Meanwhile, the second line describes the effect of the environment modeled by non-unitary interactions between replicas on forward and backward contours.

The next step is to integrate over disorder and express the path integral in terms of the Green's function $G(u,u^\prime)=\frac{1}{N}\sum_i\chi_i(u)\chi_i(u^{\prime})$ and the corresponding Lagrange multiplier, $\Sigma(u,u^\prime)$ \cite{maldacena2016}. After performing the average over random couplings, standard SYK techniques lead to $\langle\mathrm{Tr}[\mathcal{P}(t)^{m}] \rangle_{\text{dis}}=\int \mathcal{D}G\mathcal{D}\Sigma~e^{-S^{(m)}_{\mathrm{eff}}[G,\Sigma]}$ \cite{SM}. In the large $N$ limit, the saddle point approximation can be utilized. The final expression for the large $N$ second Rényi entropy is computed from the on-shell $G$-$\Sigma$ action $S = S^{(2)}_{\mathrm{eff}}[\mathcal{G},\Upsigma]-2S^{(1)}_{\mathrm{eff}}[\mathcal{G},\Upsigma]$, with $\mathcal{G}$ and $\Upsigma$ the solutions of the corresponding Schwinger-Dyson equations.

{\em \color{blue} Large $N$ Numerics.---} Except in the large $q$ limit, the saddle point equations can only be solved numerically by discretizing each time contour into $\mathcal{N}$ segments of length $\delta t$ \cite{SM}. Then, $G(u,u^\prime)$ and $\Sigma(u,u^\prime)$ are computed by solving iteratively the resulting discrete matrix equations.
\begin{figure}[t]
  \centering
  \includegraphics[width=0.84\columnwidth]{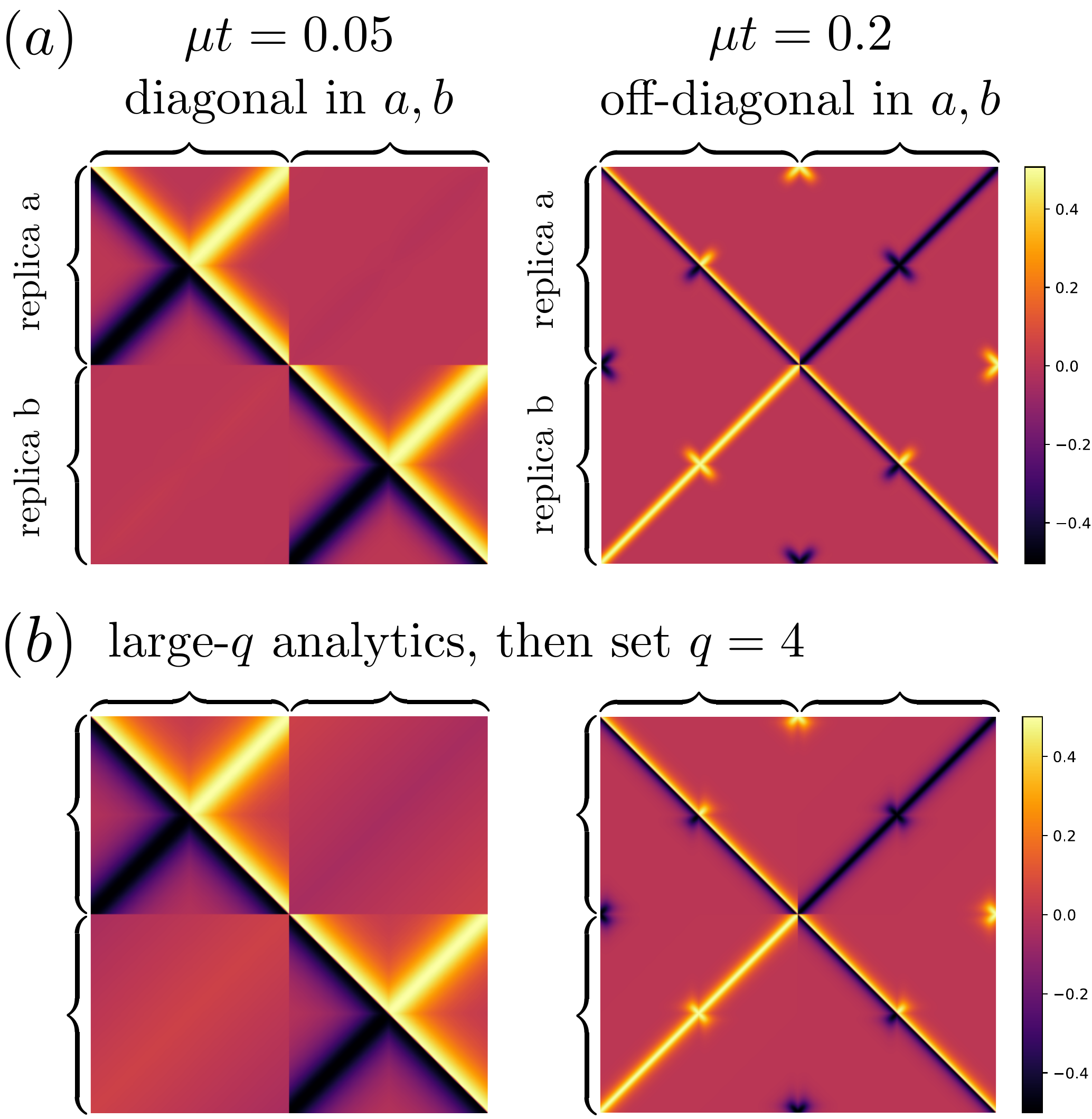}
  \caption{(a) Large $N$, $q=4$ numerical Green's function: Two types of saddle point solutions: replica-diagonal (left panel) vs replica-off-diagonal (right panel) solutions with $\mu=0.01J$. (b) Large-$q$ analytical results.}
  \label{fig:Green}
\end{figure}
\begin{figure}[t]
  \centering
  \includegraphics[width=0.7\columnwidth]{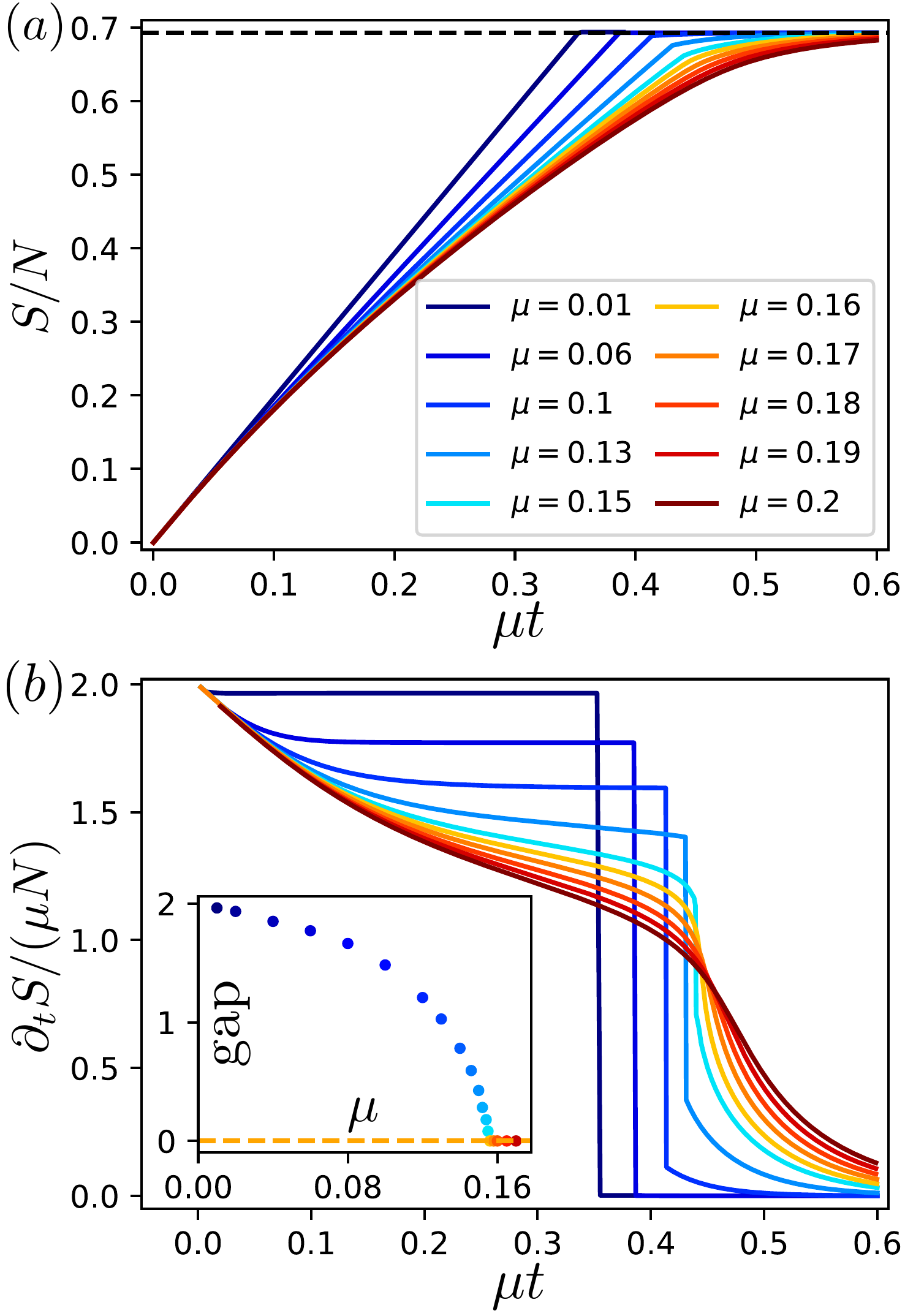}
  \caption{(a) Large $N$ Page curve for an initial $\beta J=0$ TFD state, with $\mu$ values ranging from $0.01J$ to $0.2J$. (b) Derivative of Rényi entropy respect to time, i.e., $\frac{1}{\mu N}\partial_t S$. Inset: Gap of $\frac{1}{\mu N}\partial_t S$ at Page time vs $\mu/J$. A first order phase transition at the Page time is only observed for $\mu \lesssim 0.156J$. We set $J=1$ in all numerical results.}
  \label{fig:diff_mu}
\end{figure}
For the initial state with $\beta J=0$, there exist two distinct types of saddle point solutions for $G$: the replica diagonal in the $a,b$ space (left panel of Fig.~\ref{fig:Green}(a)), and the replica off-diagonal configuration (right panel of Fig.~\ref{fig:Green}(a)). At short times and small $\mu/J$, only the replica diagonal solution is present. 
In this region,  $G^{ab} \approx G_0 \propto \delta^{a,b}$ and we find that the Rényi entropy is given by $2\mu N t$. The off-diagonal components $\sim \mu t$, see left panel of Fig.~\ref{fig:Green} are subleading at this moment. This solution cannot dominate for $t\gg 1/\mu$ because the entropy would grow indefinitely leading to the equivalent issue of the information paradox in gravity \cite{Almheiri:2019yqk}. 
Indeed, we have identified, see Fig.~\ref{fig:diff_mu}, a Page time ($\mu t \sim \mathcal{O}(1)$), at which the replica off-diagonal, in the $a,b$ space, solution becomes dominant. 
This switch of saddles at the Page time prevents the entropy from growing forever. Instead, as shown in Fig.~\ref{fig:diff_mu}(a), it saturates at $N\ln2$. The short and long time phases are clearly separated by a first-order entanglement transition in the small $\mu/J$ regime. This late time saddle is very similar to the replica wormhole reported in gravity \cite{penington2020,almheiri2020}. 

On the other hand, the transition turns into a crossover for large $\mu/J$ values. This phenomenon can be attributed to the absence of a significant symmetry difference between the two phases. For small $\mu/J$, it is possible to distinguish the two phases based on the magnitude of the off-diagonal components, which are either large, of order $\mathcal{O}(1)$, or small, of order $\mathcal{O}(\mu/J)$. However, for large $\mu/J$, the off-diagonal components no longer exhibit such a difference, leading to a hybridized and smooth behavior that results in a crossover instead of a sharp transition. By examining the derivative of the Rényi entropy, it becomes apparent that there is a noticeable gap when $\mu < \mu_c \approx 0.156J$, and the gap closes when $\mu > \mu_c$, see Fig.~\ref{fig:diff_mu}(b) and its inset. These observations hold true for the large $\beta J$ case as well, see Supplemental Material \cite{SM} for a detailed account.

{\em \color{blue} Large-$q$ Analytical Solution.---} In the large-$q$ limit, it is possible to compute Green's functions analytically \cite{maldacena2016}. We define scaled couplings $\mathcal{J}^2 = qJ^2/2^{q-1}$, $\hat{\mu}=q\mu$ which are fixed in the large $q$ limit. Green's functions are denoted by $G_{\eta\eta^\prime}^{aa} \equiv G(u_\eta^a,u_{\eta^\prime}^a)$ and $G_{\eta\eta^\prime}^{ab}\equiv G(u_\eta^a,u_{\eta^\prime}^b)$, where $\eta,\eta^\prime$ stands for forward/backward $\pm$. The difference between the two time argument is $\Delta u \equiv u-u^\prime$. For simplicity, we focus again on the $\beta J= 0$ case.

At short times $\hat{\mu} t \ll q$, the contribution from finite $\mu$ is perturbative. To leading order in  $\hat{\mu}/\mathcal{J}$, the Green's function within each replica can be approximated by the equilibrium solution with $\hat{\mu}=0$:
$G^{aa}_{\eta\eta'} =\frac{G_{0,\eta\eta'}(\Delta u)}{\cosh^{2/q}( \mathcal{J}\Delta u)}.$
Inter-replica Green's functions involves the convolution of two $G^{aa}_{\eta\eta'}$s and to lowest order it gives,
\begin{equation}\label{eq:largeq_short}
G^{ba}_{\eta\eta'}=\frac{\hat{\mu}}{2q}
\begin{pmatrix}
u+u'-2t&2t-|\Delta u|\\
2t-|\Delta u|&u+u'-2t
\end{pmatrix},
\end{equation}
which becomes large enough at sufficiently long time $t\sim q/\hat{\mu}$, where perturbation theory breakdown. By computing the on-shell action, it is straightforward to show that Eq.~\eqref{eq:largeq_short} results in $S \approx 2\hat{\mu} N t/q$, consistent with the previous perturbative result. 

On the other hand, in the long-time limit $\hat{\mu}t\rightarrow \infty$, the pairing between branches changes and the solution is replica off-diagonal. The Green's functions can be approximated by the ``factorization of twist operators'' \cite{chen2020replica}. We first consider the Greens' functions where two fermion operators are inserted in branches $\mathcal{C}_{\text{red}}=(b,+)\cup (a,-)$:
\begin{equation}
\begin{tikzpicture}[thick,scale = 0.3,baseline={([yshift=-4pt]current bounding box.center)}]
 \draw[black] (-2.5,2) arc(-90:-270:0.5 and 0.5);
 \draw[black] (2.5,2) arc(-90:90:0.5 and 0.5);

 \draw[blue]  (-2.5,2)  --  (2.5,2);
 \draw[blue]  (-2.5,3)  --  (2.5,3);

  \draw[dotted]  (2.0,0)  --  (2.0,2);
  \draw[dotted]  (1.5,0)  --  (1.5,2);
  \draw[dotted]  (1,0)  --  (1,2);
  \draw[dotted]  (0.5,0)  --  (0.5,2);
  \draw[dotted]  (0,0)  --  (0,2);
  \draw[dotted]  (-2.0,0)  --  (-2.0,2);
  \draw[dotted]  (-1.5,0)  --  (-1.5,2);
  \draw[dotted]  (-1,0)  --  (-1,2);
  \draw[dotted]  (-0.5,0)  --  (-0.5,2);

  \draw[dotted]  (2.0,4)  --  (2.0,3);
  \draw[dotted]  (1.5,4)  --  (1.5,3);
  \draw[dotted]  (1,4)  --  (1,3);
  \draw[dotted]  (0.5,4)  --  (0.5,3);
  \draw[dotted]  (0,4)  --  (0,3);
  \draw[dotted]  (-2.0,4)  --  (-2.0,3);
  \draw[dotted]  (-1.5,4)  --  (-1.5,3);
  \draw[dotted]  (-1,4)  --  (-1,3);
  \draw[dotted]  (-0.5,4)  --  (-0.5,3);

  \draw[dotted]  (2.0,-2)  --  (2.0,-1);
  \draw[dotted]  (1.5,-2)  --  (1.5,-1);
  \draw[dotted]  (1,-2)  --  (1,-1);
  \draw[dotted]  (0.5,-2)  --  (0.5,-1);
  \draw[dotted]  (0,-2)  --  (0,-1);
  \draw[dotted]  (-2.0,-2)  --  (-2.0,-1);
  \draw[dotted]  (-1.5,-2)  --  (-1.5,-1);
  \draw[dotted]  (-1,-2)  --  (-1,-1);
  \draw[dotted]  (-0.5,-2)  --  (-0.5,-1);

  \draw[mid arrow, blue] (2.5,3) -- (-2.5,3);
  \draw[mid arrow, red] (-2.5,2) -- (2.5,2);

 \draw[black] (-2.5,2-3) arc(-90:-270:0.5 and 0.5);
 \draw[black] (2.5,2-3) arc(-90:90:0.5 and 0.5);

 \draw[blue]  (-2.5,2-3)  --  (2.5,2-3);
 \draw[blue]  (-2.5,3-3)  --  (2.5,3-3);

  \draw[mid arrow, red] (2.5,3-3) -- (-2.5,3-3);
  \draw[mid arrow, blue] (-2.5,2-3) -- (2.5,2-3);
  \draw[orange,dashed]  (-3.5,4.5)  --  (-3.5,-2.5);
  \draw[orange,dashed]  (1,4.5)  --  (1,-2.5);
  \draw[orange,dashed]  (-3.5,4.5)  --  (1,4.5);
  \draw[orange,dashed]  (-3.5,-2.5)  --  (1,-2.5);
\filldraw  (-1.4,2) circle (2pt) node[left]{\scriptsize $ $};
\filldraw  (-0.8,0) circle (2pt) node[left]{\scriptsize $ $};

\end{tikzpicture}
\ \ \approx\ \
\begin{tikzpicture}[thick,scale = 0.3,baseline={([yshift=-4pt]current bounding box.center)}]
 \draw[black] (-2.5,2) arc(-90:-270:0.5 and 0.5);
 \draw[black] (2.5,3) arc(90:0:0.5 and 0.5);

 \draw[blue]  (-2.5,2)  --  (2.5,2);
 \draw[blue]  (-2.5,3)  --  (2.5,3);

  \draw[dotted]  (2.0,0)  --  (2.0,2);
  \draw[dotted]  (1.5,0)  --  (1.5,2);
  \draw[dotted]  (1,0)  --  (1,2);
  \draw[dotted]  (0.5,0)  --  (0.5,2);
  \draw[dotted]  (0,0)  --  (0,2);
  \draw[dotted]  (-2.0,0)  --  (-2.0,2);
  \draw[dotted]  (-1.5,0)  --  (-1.5,2);
  \draw[dotted]  (-1,0)  --  (-1,2);
  \draw[dotted]  (-0.5,0)  --  (-0.5,2);

  \draw[dotted]  (2.0,4)  --  (2.0,3);
  \draw[dotted]  (1.5,4)  --  (1.5,3);
  \draw[dotted]  (1,4)  --  (1,3);
  \draw[dotted]  (0.5,4)  --  (0.5,3);
  \draw[dotted]  (0,4)  --  (0,3);
  \draw[dotted]  (-2.0,4)  --  (-2.0,3);
  \draw[dotted]  (-1.5,4)  --  (-1.5,3);
  \draw[dotted]  (-1,4)  --  (-1,3);
  \draw[dotted]  (-0.5,4)  --  (-0.5,3);

  \draw[dotted]  (2.0,-2)  --  (2.0,-1);
  \draw[dotted]  (1.5,-2)  --  (1.5,-1);
  \draw[dotted]  (1,-2)  --  (1,-1);
  \draw[dotted]  (0.5,-2)  --  (0.5,-1);
  \draw[dotted]  (0,-2)  --  (0,-1);
  \draw[dotted]  (-2.0,-2)  --  (-2.0,-1);
  \draw[dotted]  (-1.5,-2)  --  (-1.5,-1);
  \draw[dotted]  (-1,-2)  --  (-1,-1);
  \draw[dotted]  (-0.5,-2)  --  (-0.5,-1);

  \draw[mid arrow, blue] (2.5,3) -- (-2.5,3);
  \draw[mid arrow, red] (-2.5,2) -- (2.5,2);

\draw[black] (3,-0.5) -- (3,2.5);
\draw[black] (2.5,-0) -- (2.5,2);

 \draw[black] (-2.5,2-3) arc(-90:-270:0.5 and 0.5);
 \draw[black] (2.5,2-3) arc(-90:0:0.5 and 0.5);

 \draw[blue]  (-2.5,2-3)  --  (2.5,2-3);
 \draw[blue]  (-2.5,3-3)  --  (2.5,3-3);

  \draw[mid arrow, red] (2.5,3-3) -- (-2.5,3-3);
  \draw[mid arrow, blue] (-2.5,2-3) -- (2.5,2-3);

  \draw[orange,dashed]  (-3.5,4.5)  --  (-3.5,-2.5);
  \draw[orange,dashed]  (1,4.5)  --  (1,-2.5);
  \draw[orange,dashed]  (-3.5,4.5)  --  (1,4.5);
  \draw[orange,dashed]  (-3.5,-2.5)  --  (1,-2.5);
\filldraw  (-1.4,2) circle (2pt) node[left]{\scriptsize $ $};
\filldraw  (-0.8,0) circle (2pt) node[left]{\scriptsize $ $};

\end{tikzpicture}
\ \ = \ \ 
\begin{tikzpicture}[thick,scale = 0.3,baseline={([yshift=-4pt]current bounding box.center)}]

  \draw[dotted]  (2.0,0)  --  (2.0,2);
  \draw[dotted]  (1.5,0)  --  (1.5,2);
  \draw[dotted]  (1,0)  --  (1,2);
  \draw[dotted]  (0.5,0)  --  (0.5,2);
  \draw[dotted]  (0,0)  --  (0,2);
  \draw[dotted]  (-2.0,0)  --  (-2.0,2);
  \draw[dotted]  (-1.5,0)  --  (-1.5,2);
  \draw[dotted]  (-1,0)  --  (-1,2);
  \draw[dotted]  (-0.5,0)  --  (-0.5,2);

  \draw[mid arrow, red] (-2.5,2) -- (2.5,2);

\draw[black] (2.5,-0) -- (2.5,2);
\draw[black] (-2.5,-0) -- (-2.5,2);

 \draw[blue]  (-2.5,3-3)  --  (2.5,3-3);

  \draw[mid arrow, red] (2.5,3-3) -- (-2.5,3-3);

\filldraw  (-1.4,2) circle (2pt) node[left]{\scriptsize $ $};
\filldraw  (-0.8,0) circle (2pt) node[left]{\scriptsize $ $};
\end{tikzpicture}
\label{factorization}
\end{equation}
Here, the black dots represent the insertion of Majorana operators and the orange dashed box indicates the region in which Green's functions are the same. Green's functions on the R.H.S. exactly match those on the traditional Keldysh contour for the evolution of the density matrix of the single-sided steady state $\rho=2^{-N/2}\mathbb{I}$ \cite{kulkarni2022,garcia2022e,kawabata2022} 

Consequently, the Green's function on $\mathcal{C}_\text{red}$ or $\mathcal{C}_\text{blue}$ matches the equilibrium Green's functions with $G^{bb}_{++}=-G^{aa}_{--}=\text{sgn}(\Delta u)G^{ba}_{+-}=-\text{sgn}(\Delta u)G^{ab}_{-+}$, and
\begin{equation}
\begin{aligned}
&G^{ba}_{+-}=\frac{1}{2}\left[\frac{A}{\cosh(B+A \mathcal{J}|\Delta u|)}\right]^{2/q},
\end{aligned}
\end{equation}
where $A=\cosh B$ and $\hat{\mu}=2\mathcal{J}\sinh B$ are determined from the boundary conditions \cite{SM}.

The remaining task involves computing the correlation between $\mathcal{C}_\text{red}$ and $\mathcal{C}_\text{blue}$ for long times. It is expected that this correlation will be localized around the boundary twists. As a representative example, we  evaluate $G^{aa}_{-+}$, see Eq.~\eqref{factorization}, with the understanding that other components can be derived by symmetry. Expanding $G^{aa}_{-+}=\frac{1}{2}(1+g^{aa}_{-+}/q+...)$, one arrives at the Liouville equation $\partial_{u}\partial_{u'}g^{aa}_{-+}=2\mathcal{J}^2 e^{g^{aa}_{-+}}.$
This equation is solved, see Supplemental Material \cite{SM}, for the relevant boundary conditions. The resulting large-$q$ analytical Green's function reads
\begin{myequation}\label{eq:large_q_sol_long}
G^{aa}_{-+}=\frac{1}{2}\left[\frac{A \,\text{csch}(A \mathcal{J}u)\, \text{csch}(A \mathcal{J}u'+B)}{\coth (A \mathcal{J}u'+B)
   [\coth (A \mathcal{J}u)+2 \tanh B]-1}\right]^{2/q}.\notag
\end{myequation}
This is depicted in Fig.~\ref{fig:Green}(b), which shows a good agreement with the $q=4$ numerical result. Furthermore, using the large-$q$ solution, we show that the long-time entropy $S$ is independent of $\mathcal{J}$ or $\hat{\mu}$ by computing $\partial_\mathcal{J} S=\partial_{\hat{\mu}} S=0$ \cite{SM}. On the other hand, we expect $S=N\ln 2$ for $\hat{\mu}/\mathcal{J}\rightarrow \infty$ where the coupling to the bath dominates. We thus conclude $S=N\ln 2$ in the long-time limit for arbitrary $\mathcal{J}/\hat{\mu}$. Although we cannot obtain analytic results for intermediate times where the transition occurs, the simple extrapolation of the short and long time analytic results to intermediate times yields a time evolution of the purity very close to the large $N$ numerical results obtained in the previous section.

{\em \color{blue} Finite $N$ Trajectories.---}%
We now investigate whether signatures of replica wormholes are visible  for finite $N$. The standard approach of duplicating the degrees of freedom of the original system using the Choi-Jamiolkowski isomorphism
\cite{2003JPhA...3610101T,PhysRevLett.93.207205,2008NJPh...10d3026P,zhou2021renyi} is not viable in our case due to the presence of four time contours, which limits exact diagonalization up to $N \approx 6$. Instead, we exploit the Markovian properties of the Lindblad coupling which allows to evolve only a single SYK. This is achieved by introducing {\em quantum trajectories} \cite{molmer1993monte,wiseman1996quantum}, which ensures that the coupling to the bath is considered even when replicas $a$ and $b$ are decoupled, see Fig.~\ref{fig:setup}(d). 

Specifically, we discretize time into small intervals with length $\delta t$. The evolution is determined by a random variable $\xi(u)\in(0,1)$ at time $u$,
\begin{align}\label{eq:infinitesimal}
\mathcal{U}_{\pm}(\delta t,\xi(u)) = &
  \begin{cases}
     e^{\mp\mathrm{i}H\delta t} &\text{, if }\xi(u) > 1-\delta p \\
     \sqrt{2}\chi_i &\text{, if }\xi(u) < 1-\delta p\\
  \end{cases}
\end{align}
where the leakage probability is given by $\delta p = 1-e^{-\mu N\delta t/2}$, indicating the probability of being detected and measured by the environment, i.e., experiencing a quantum jump. $\boldsymbol{\xi} = \{\xi(\delta t), \,\xi(2\delta t),...,\,\xi({\mathcal{N}_t}\cdot\delta t)\}$ then defines the full trajectory that determines the evolution operator and therefore the quantum dynamics. Subsequently, $\mathrm{Tr}(\mathcal{P}^2)$ is computed by substituting the connecting dots between forward/backward contours in Fig.~\ref{fig:setup}(c) with identical, yet time-reversed, quantum trajectories. See Fig.~\ref{fig:setup}(d) for a pictorial demonstration. \\
After averaging over all trajectories, we arrive at $\mathrm{Tr}(\mathcal{P}^2) = \overline{\mathcal{A}\cdot \mathcal{A}^*}$, where $\mathcal{A}$ denotes the amplitude in replica $a$, $\mathcal{A}^*$ is its time-reversed counterpart in replica $b$, and
\begin{equation}
\begin{aligned}
\mathcal{A} = \mathrm{Tr}\Big[&\mathcal{U}_+(t,\boldsymbol{\xi}_\text{L}^{+})e^{-\beta H/2}\mathcal{U}_+(t,\boldsymbol{\xi}_\text{R}^{+}) \\
\cdot\, &\mathcal{U}_-(t,\boldsymbol{\xi}_\text{R}^{-})e^{-\beta H/2}\mathcal{U}_-(t,\boldsymbol{\xi}_\text{L}^{-})\Big].
\end{aligned}
\end{equation}
After normalization, the purity is determined using the above expression after averaging, denoted by $\overline{\gamma}$, over all trajectories.
\begin{figure}[t]
  \centering
  \includegraphics[width=0.83\columnwidth]{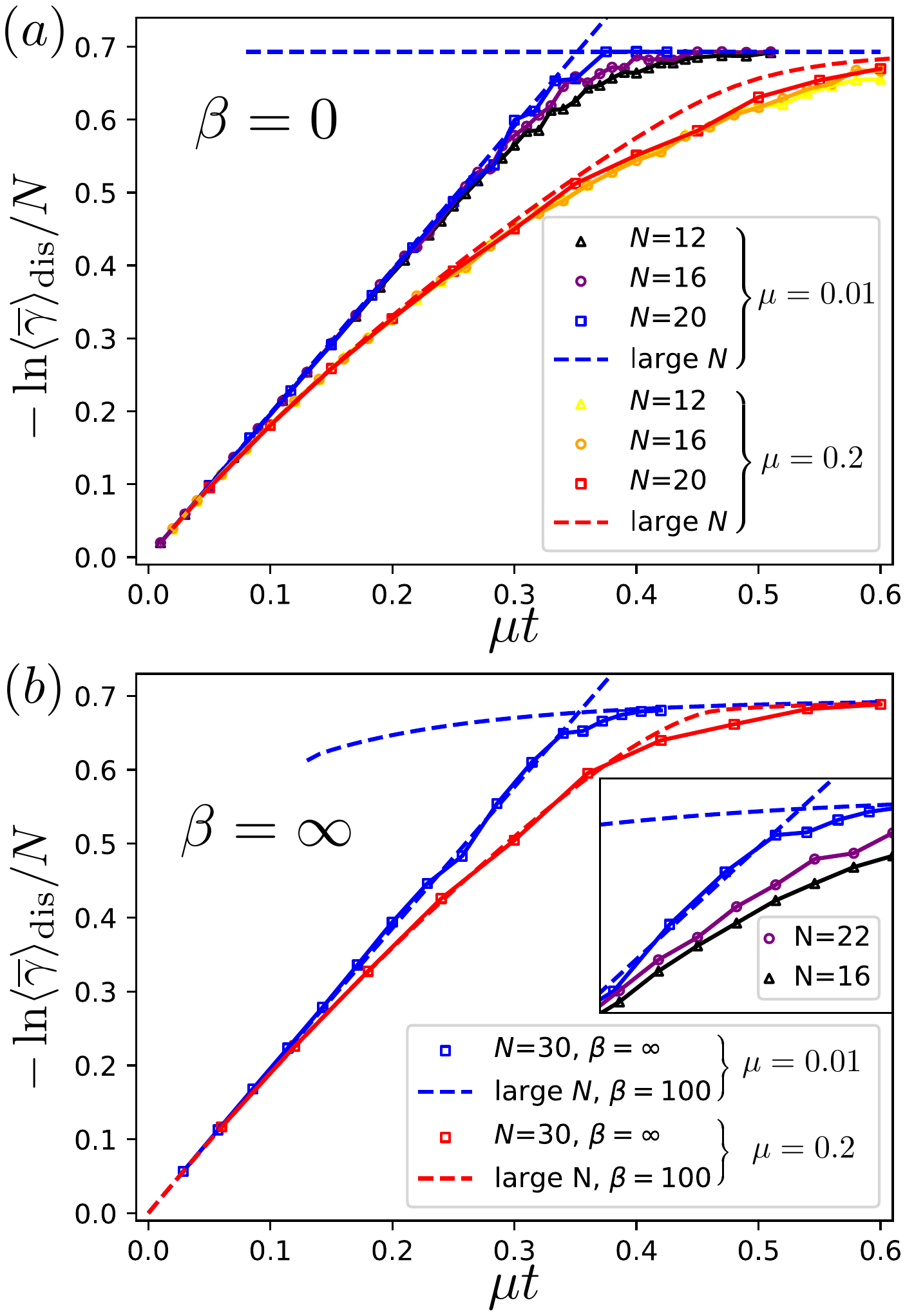}
  \caption{Comparison between the numerical entropy growth at finite $N$, computed by performing an average over quantum trajectories, and the previous large $N$ results. We present results for (a) $\beta J = 0$ and (b) $\beta J= \infty$. Inset of (b) refers to the case of $N=16,22,30$ at $\mu=0.01J$, emphasized on the transition region.}
  \label{fig:seperation}
\end{figure}

We employ the Monte Carlo method to simulate \cite{garcia2023prepare,Liu2023} the infinitesimal random evolution, Eq.~(\ref{eq:infinitesimal}).
This formalism makes possible the calculation of $\overline{\gamma}$ at $\beta J=0$ and $\beta J=\infty$ up to  $N = 20$ and $N = 30$, respectively. 
The results for the entropy growth, shown in Fig.~\ref{fig:seperation} for various values of $\mu$, exhibit excellent agreement with the large $N$ calculation for the largest $N$ we can explore numerically. We stress that for this largest $N$, the numerical results captures both the change in leading saddles for small $\mu$ and the termination of the first-order transition for $\mu \gtrsim \mu_c$. However, for smaller values of $N$, significant differences for intermediate times make it challenging to distinguish the transition from the crossover as $\mu$ increases.
This observation confirms the crucial role of the quantum trajectory method, achieving a dramatic, up to five-fold, increase in system size, in order to reproduce the large $N$ results. Our numerics confirms the possibility to observe entanglement dynamics with features of replica wormholes in quantum systems with moderate $N$ which makes more likely its experimental confirmation.

{\em \color{blue} Conclusions.---}
In this work, we investigate the entanglement growth of a pair of dissipative SYK models. In the large $N$ limit, we discover a first-order entanglement transition for small dissipation strengths. Notably, we observe that the long-time saddle closely resembles replica wormholes in the context of gravity. We further conduct numerical calculations at finite $N$ using the quantum trajectories method. The results clearly demonstrate that the signature of the replica wormhole remains visible even at moderate $N$. We expect that our findings can be tested using state-of-the-art platforms for quantum simulations \cite{Jafferis:2022crx} or experimentally in strongly interacting open systems, thus offering a novel playground to observe wormholes \cite{PhysRevX.12.031013,PRXQuantum.4.010320,PRXQuantum.4.010321} by their role on entanglement properties \cite{huang2020predicting,elben2020cross,rath2021importance}.

\begin{acknowledgments}
We are grateful to A. Kamenev for useful discussions. HW, CL, and AMGG acknowledge support from a National Key R\&D Program of China (Project ID: 2019YFA0308603). HW is partially supported by NSFC Grant No.~12247131. AMGG also acknowledges partial support from a Shanghai talent program.
\end{acknowledgments}


\bibliography{SYK.bib}

\ifarXiv
    

\section*{Supplementary material}

# Supplemental Material: Entanglement Transition and Replica Wormhole in the Dissipative Sachdev-Ye-Kitaev Model

Hanteng Wang,[1, *] Chang Liu,[1, †] Pengfei Zhang,[2, ‡] and Antonio M. García-García[1, §]

[1]*Shanghai Center for Complex Physics, School of Physics and Astronomy,*
*Shanghai Jiao Tong University, Shanghai 200240, China*
[2]*Department of Physics, Fudan University, Shanghai, 200438, China*

In this Supplemental Material, we provide addition details about (I) the derivation of the Keldysh-Lindblad path integral for the purity and (II) the analytical large-$q$ calculation.

## I. KELDYSH-LINDBLAD PATH INTEGRAL FOR PURITY

### A. Contours and actions

The unitary evolution of the L-R system is governed by the Hamiltonian $H_\text{S} = H_\text{L} + H_\text{R}$. For a given time $t$, the wavefunction for the double-sided SYK system evolves according to

$$|\Psi(t)\rangle = e^{-iH_\text{S}t}|\text{TFD}_\beta\rangle = Z_\beta^{1/2} e^{-i(H_\text{L}+H_\text{R})t} e^{-\beta(H_\text{L}+H_\text{R})/4}|I\rangle, \quad (1)$$

where the Majorana fermion on each side of the system satisfies $\chi_{i,\text{L}}|I\rangle = -i\chi_{i,\text{R}}|I\rangle$. Under this convention, it's possible to map the double-sided SYK back to a single-sided SYK. The wavefunction, projected onto a Majorana coherent state, can be derived and represented as follows [1]:

$$\langle \chi_\text{L}, \chi_\text{R}|\Psi(t)\rangle \;=\; \underset{e^{-iH_\text{S}t}}{\text{TFD}} \xrightarrow{\text{map to single-sided SYK}} \chi_\text{L} \underset{t\quad\quad t}{\overset{\frac{1}{2}\beta}{\frown}} \chi_\text{R} \quad (2)$$

In the single-sided SYK picture, the L-R forward evolution are denoted by two left-pointing arrow lines $U_+ = e^{-iHt}$ separated by a semi-circle that represents the imaginary evolution $U_{\beta/2} = e^{-\beta H/2}$. In the absence of any Lindbladian effects, one can define the corresponding unnormalized density matrix as $\mathcal{P}(t) = Z_\beta|\Psi(t)\rangle\langle\Psi(t)|$. Taking into account the conjugate part, one needs the backward evolution $U_- = e^{iHt}$ which is represented by a right-pointing arrow line. Thus one can express the density matrix via the matrix element as:

$$\langle \chi_\text{L}, \chi_\text{R}|\mathcal{P}(t)|\chi'_\text{L}, \chi'_\text{R}\rangle \;=\; \begin{matrix}\chi_\text{L} & & \chi_\text{R}\\ & \frown & \\ \chi'_\text{L} & & \chi'_\text{R}\end{matrix} \quad (3)$$

Here, the matrix element comprises four external legs: $\chi_\text{L}, \chi_\text{R}$ as initial legs, and $\chi'_\text{L}, \chi'_\text{R}$ as final legs. This construction is distinct from the case of the one-sided SYK initialized in a thermal state, where the density matrix only contains two external legs [2].

When a Markovian environment is incorporated, as modeled by jump operators in the Lindblad formalism, it results in an effective coupling in the path integral between forward and backward contours at equal time. This originates from the left and right sides (which should not be confused with the L-R of the SYK systems, and should be denoted as $\pm$) of the density matrix $\mathcal{P}$. Within the infinitesimal time $\delta t$, the non-unitary equation of motion can be mapped to the path integral, provided the single-sided jump operators $L_i$ are defined on the forward/backward contours as $L_{i,\pm}$. After integrating over a finite time on the contour of Eq. (3) and identifying $L_{i,\pm}(s) = \chi_i(s_\pm)$, we attain the following contribution for the density matrix [3–7]:

$$\delta\mathcal{P}/\delta t = \mu \sum_i \left[L_i\mathcal{P}L_i^\dagger - \frac{1}{2}\{L_i^\dagger L_i, \mathcal{P}\}\right] \to \exp\left[\int \delta t\, \mu \sum_i \left(L_{i,+}L_{i,-}^\dagger - \frac{1}{2}L_{i,+}^\dagger L_{i,+} - \frac{1}{2}L_{i,-}^\dagger L_{i,-}\right)\right]$$

$$= \exp\left[\int ds\, \mu \sum_i \chi_i(s_+)\chi_i(s_-) - \mu N t\right] \quad (4)$$

Here we use a time argument $s$ to denote evolution and $s_+/s_-$ are equal time on forward/backward branch. As a result, we can illustrate the density evolution of L-R systems with a Markovian reservoir as:

$$\mathcal{P}(t) = \quad \begin{array}{c} s_+ \\ s_- \end{array} \tag{5}$$

where the vertical dotted lines refers to the Lindblad couplings between forward/backward contours at equal time.

With knowledge of the density matrix at time $t$, the unnormalized purity $\text{Tr}[\mathcal{P}(t)^2]$ can now be formulated by introducing two replicas, namely 1 and 2. More specifically, in the basis of the Majorana coherent state, we can express the purity by linking the legs of two replicas so that the procedure is consistent with the operation of taking the trace [8, 9],

$$\text{Tr}[\mathcal{P} \cdot \mathcal{P}] = \sum_{\chi \text{ in proper measure}} \overbrace{\langle \chi_\text{L}, \chi_\text{R}|\mathcal{P}(t)|\chi'_\text{L}, \chi'_\text{R}\rangle}^{\text{replica 1}} \underbrace{\langle \chi'_\text{L}, \chi'_\text{R}|\mathcal{P}(t)|\chi_\text{L}, \chi_\text{R}\rangle}_{\text{replica 2}} = \left[\begin{array}{c} s^1_+ \\ s^1_- \\ s^2_+ \\ s^2_- \end{array}\right] \tag{6}$$

The TFD state leads to entanglement between L-R systems. Hence, calculating the purity of this state results in a twisted link between the legs, which we depict in the diagram above. The presence of this twisted boundary condition between replicas 1 and 2, complicates the calculation of the purity. In particular, the definition of $\partial_s$, which typically arises in the path integral form $\chi(s)\partial_s\chi(s)$ to encode time ordering, becomes subtle to formulate at the twisting positions. Fortunately, there is a workaround that allows to fix this issue. By reshuffling the time contour, we can redefine new replicas, denoted by $a$ and $b$. This process is illustrated in the following diagram:

$$\text{Tr}[\mathcal{P}(t)^2] = \quad \begin{array}{c} u^a_+ \\ u^a_- \\ u^b_+ \\ u^b_- \end{array} \tag{7}$$

where the time domain along the contour $\mathcal{C}$ is denoted as $u \in [0, 8t+4\beta]$, with the parametrization of

$$\text{forward}: \begin{cases} u^a_+ = u \in [0, t] \cup [t + \tfrac{1}{2}\beta, 2t + \tfrac{1}{2}\beta] \\ u^b_+ = u \in [4t+\beta, 5t+\beta] \cup [5t + \tfrac{3}{2}\beta, 6t + \tfrac{3}{2}\beta] \end{cases}$$

$$\text{backward}: \begin{cases} u^a_- = u \in [2t + \tfrac{1}{2}\beta, 3t + \tfrac{1}{2}\beta] \cup [3t+\beta, 4t+\beta] \\ u^b_- = u \in [6t + \tfrac{3}{2}\beta, 7t + \tfrac{3}{2}\beta] \cup [7t+2\beta, 8t+2\beta] \end{cases}$$

$$\text{imaginary}: u \in [t, t+\tfrac{1}{2}\beta] \cup [3t+\tfrac{1}{2}\beta, 3t+\beta] \cup [5t+\beta, 5t+\tfrac{3}{2}\beta] \cup [7t+\tfrac{3}{2}\beta, 7t+2\beta].$$

Given the time domain $u$ and the corresponding replicas $a, b$, it's straightforward to determine the time derivative operation $\partial_u$ with respect to $u$. In the space of $a, b$, this operation is diagonal, represented as $\partial_u = G_0^{-1}$, where $G_0^{aa} = G_0^{bb} = \tfrac{1}{2}\text{sgn}(u - u')$ and $G_0^{ab} = G_0^{ba} = 0$. This $\partial_u$ also imposes the fermionic anti-periodic boundary condition in each replica automatically. The translation of this diagram Eq. (7) back into the path integral representing the purity results in,

$$\text{Tr}(\mathcal{P}^2) = \int \mathcal{D}\chi \exp\left\{-\int_\mathcal{C} du \left[\frac{1}{2}\sum_i^N \chi_i \partial_u \chi_i + f(u)H\right] + \mu \sum_i^N \int du \left[\chi_i(u^a_+)\chi_i(u^b_-) + \chi_i(u^a_-)\chi_i(u^b_+)\right] - 2\mu N t\right\}. \tag{8}$$

The next step is to perform an average over disorder realizations, namely the SYK random couplings, $\langle \text{Tr}(\mathcal{P}^2)\rangle_\text{dis}$ and the definition of the Green's function $G(u, u') = \frac{1}{N}\sum_i \chi_i(u)\chi_i(u')$ and the corresponding Lagrange multiplier,

$\Sigma(u, u')$. This results in the following $m^{\text{th}}$-effective action of $\langle \text{Tr}(\mathcal{P}^m) \rangle_{\text{dis}}$,

$$S_{\text{eff}}^{(m)}[G, \Sigma] = -\frac{N}{2} \log \det(\partial_u - \Sigma) + \frac{N}{2} \int_{\mathcal{C}} du du' \left[ \Sigma G - f(u)f(u') \frac{J^2}{q} G^q \right] - \frac{\mu N}{2} \int_{\mathcal{C}} du du' G(u, u') g(u, u') + m \mu N t, \tag{9}$$

where $f(u)$ and $g(u, u')$ stand for,

$$f(u) = \begin{cases} i & u \in \text{forward} \\ -i & u \in \text{backward} \\ 1 & u \in \text{imaginary} \end{cases} \quad \& \quad g(u, u') = \begin{cases} +\delta(8t + 2\beta - u - u') & u \in \text{forward} \\ -\delta(8t + 2\beta - u - u') & u \in \text{backward} \end{cases} \tag{10}$$

For $m = 2$, and in large $N$ limit, we compute the path integral in the saddle point approximation leading to $\text{Tr}(\mathcal{P}^2) \simeq e^{-S_{\text{eff}}^{(2)}[\mathcal{G}, \Sigma]}$, where $S_{\text{eff}}^{(2)}[\mathcal{G}, \Sigma]$ is the on-shell action whose arguments follow the saddle points, usually termed Schwinger-Dyson, equations

$$\begin{aligned} \mathcal{G}(u, u') &= [\partial_u - \Sigma(u, u')]^{-1} \\ \Sigma(u, u') &= J^2 \mathcal{G}(u, u')^q f(u)f(u') + \mu g(u, u') \end{aligned} \tag{11}$$

The normalization can be calculated by using the $m = 1$ single replica action, which eventually yields $Z_\beta = \text{Tr}(\mathcal{P}) \simeq e^{-S_{\text{eff}}^{(1)}}$. Finally, the Renyi entropy in large $N$ limit becomes $S = S_{\text{eff}}^{(2)}[\mathcal{G}, \Sigma] - 2 S_{\text{eff}}^{(1)}[\mathcal{G}, \Sigma]$.

## B. Solutions

The Schwinger-Dyson equations above are solved iteratively once the quantities $G$ and $\Sigma$ are discretized into a matrix form [10]. These equations are represented in two different replica spaces, $a \& b$ [9] and $1 \& 2$ [8], as illustrated in Fig. 1. The left and right panels display two distinct saddle point solutions of Eq. (11), calculated at the same time $t$. For early times, before the Page time where the transition occurs, the solution that is almost diagonal in the $a, b$ space (left panel of Fig. 1(a)) exhibits lower entropy. However, for sufficiently long times, longer than the Page time, the solution which is nearly diagonal in the 1, 2 space (right panel of Fig. 1(b)) tends to have a lower entropy. The physical interpretation of the process, when viewed from the 1, 2 replica space representation, can be outlined as follows: Initially, L-R pairs of the SYKs are highly entangled, but under the impact of environmental interactions, they part ways and start entangling more intensively with their respective environments at the Page time. This illuminates the inherent monogamy feature of quantum entanglement [11, 12].

We have also calculated the growth of the purity for a initial TFD state at low temperature $\beta J = 100$. As is observed in Fig. 2(b), there is no qualitative difference between initial TFD states at low and high temperature.

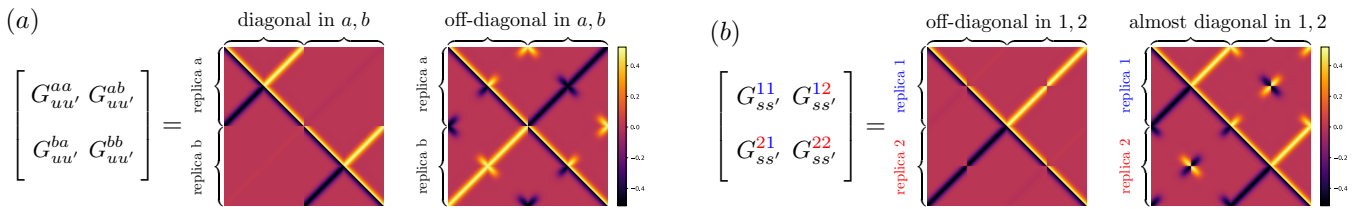

FIG. 1. Two distinct saddle point solutions for Eq. (11): replica diagonal vs. replica off-diagonal at time $t = 12/J$, with $\mu = 0.01J$ and $\beta = 0$. (a) $G(u, u')$ displayed in replica representations $a$ and $b$. The left panel illustrates the replica diagonal solution, while the right panel shows the replica off-diagonal solution at the same time. (b) $G(s, s')$ presented in replica representations 1 and 2.

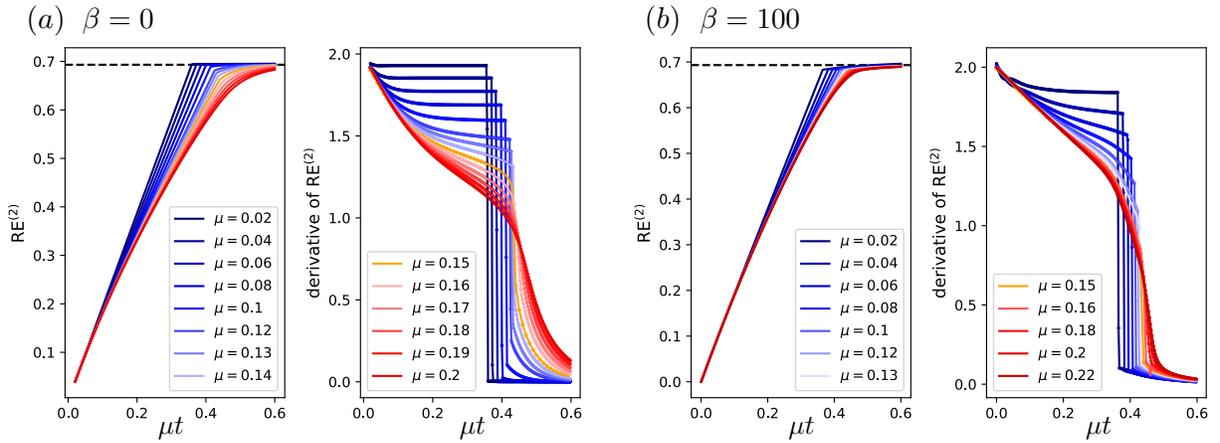

FIG. 2. (a) Left panel: Large $N$ Page curve for an initial $\beta J = 0$ TFD state, for values of the coupling to the bath $\mu$ ranging from $0.02J$ to $0.2J$. Right panel: Time dependence of the derivative of the Rényi entropy for different values of $\mu$ as a function of the rescaled time $\mu t$. (b) Large $N$ Page curve for an initial $\beta J = 100$ TFD state.

## II. THE LARGE-$q$ SYK WITH DISSIPATION

### A. Single-sided dissipative SYK

We provided additional details of analytical large $q$ calculation presented in the main text. Our starting point is a single-sided large-$q$ SYK model with Hamiltonian

$$H = \sum_{1 \leq i_1 < i_2 < \ldots < i_q \leq N} i^{q/2} J_{i_1 i_2 \ldots i_q} \chi_{i_1} \chi_{i_2} \cdots \chi_{i_q}. \tag{12}$$

$J_{i_1 i_2 \ldots i_q}$ with different set of indices are independent Gaussian variables, with expectations and variances given by

$$\langle J_{i_1 i_2 \ldots i_q} \rangle_{\text{dis}} = 0, \qquad \langle J^2_{i_1 i_2 \ldots i_q} \rangle_{\text{dis}} = \frac{(q-1)! J^2}{N^{q-1}} = \frac{2^{q-1}(q-1)! \mathcal{J}^2}{q N^{q-1}}. \tag{13}$$

We include a Markovian bath by considering the Lindblad master equation with jump operators $L_i = \sqrt{\mu} \chi_i$:

$$\partial_t \rho = -i[H, \rho] + \sum_i \mu \left( \chi_i \rho \chi_i - \frac{1}{2} \rho \right), \tag{14}$$

where $\rho$ is the density matrix. Note that we employ a different notation for the density matrix from the one used in the main text, $\mathcal{P}$ where both L-R systems are considered. This model, and its solutions, are discussed in Refs. [5–7]. Here we just summarizes these results for later use. A useful pictorial representation of the Keldysh contour at $\beta = 0$ is,

$$\begin{array}{c} u_+ \\ \overline{\phantom{xxxxxxxxx}} \\ u_- \end{array} \tag{15}$$

where the solid arrow lines represent the branch with forward/backward evolution in time. The dotted lines represent the coupling induced by the environment. We focus on the time evolution towards the steady state with $\rho = 2^{-N/2} \mathcal{I}$. For that purpose, we compute the Green's functions $G_{\eta \eta'}(u, u')$. The Schwinger-Dyson equation reads

$$\begin{pmatrix} \partial_u - \Sigma_{++} & \mu - \Sigma_{+-} \\ -\mu - \Sigma_{-+} & -\partial_u - \Sigma_{--} \end{pmatrix} \circ \begin{pmatrix} G_{++} & G_{+-} \\ G_{-+} & G_{--} \end{pmatrix} = \hat{I}.$$

$$\Sigma_{\eta \eta'} = -\eta \eta' \frac{\mathcal{J}^2}{q} (2 G_{\eta \eta'})^{q-1}. \tag{16}$$

We introduce $\mu = \hat{\mu}/q$ and fix $\hat{\mu}$ in the large-$q$ limit. At $q = \infty$, both $\mu$ and $\Sigma$ becomes zero. Consequently, the Green's functions reads

$$G_{0,++} = -G_{0,--} = \frac{1}{2}\text{sgn}(u-u') = \frac{1}{2}\text{sgn}(\Delta u), \quad G_{0,-+} = -G_{0,+-} = \frac{1}{2}. \tag{17}$$

We then expand $G_{\eta\eta'} = G_{0,\eta\eta'}(1 + g_{\eta\eta'}/q + ...)$ in the large-$q$ limit. In Majorana systems, we can focus on $G_{-+}$, which determines all other Green's functions due to the symmetry of the Keldysh contour. Keeping only the leading $1/q$ corrections, the Schwinger-Dyson equation for $G_{-+}$ becomes

$$\partial_u \partial_{u'} g_{-+} = 2\mathcal{J}^2 e^{g_{-+}} + 2\hat{\mu}\delta(\Delta u). \tag{18}$$

We first focus on $u > u'$, where (18) becomes a Liouville equation. The translational invariant solution reads

$$e^{g_{-+}} = \frac{A^2}{\cosh^2(B + A\mathcal{J}\Delta u)} \quad \text{for } \Delta u > 0. \tag{19}$$

Due to the reflection symmetry of the Keldysh contour with initial density matrix $\rho = 2^{-N/2}\mathcal{I}$, we have $g_{-+}(\Delta u) = g_{-+}(-\Delta u)$. Parameters $A$ and $B$ are determined from the boundary condition at $t = 0$

$$g(0) = 0, \qquad \partial_u g(0^+) = -\hat{\mu}. \tag{20}$$

The first boundary condition comes from $\chi^2 = 1/2$ and the second boundary condition takes into account the last term in (18) which results in

$$A = \cosh B, \qquad \hat{\mu} = 2\mathcal{J}\sinh B. \tag{21}$$

In the special limit $\hat{\mu} \to 0$, we find $B = 0$ and $A = 1$.

## B. The purity calculation

Now we turn to the purity calculation involving two copies of the density matrix investigated in the main text. For $\beta = 0$, the Keldysh contour with time $2t$ becomes

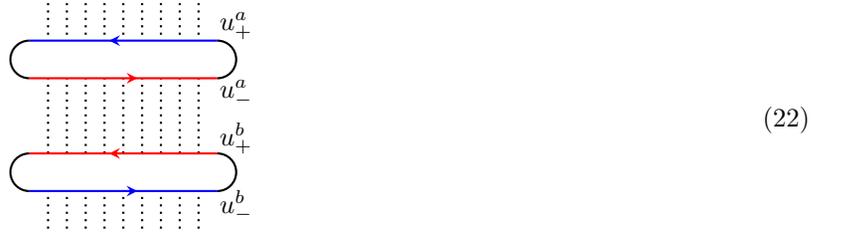

(22)

The Schwinger-Dyson equation on this doubled Keldysh contour reads:

$$\sum_{p''} \begin{pmatrix} \delta^{pp''}\partial_u - \Sigma_{++}^{pp''} & \mu\delta^{\bar{p}p'} - \Sigma_{+-}^{pp''} \\ -\mu\delta^{\bar{p}p'} - \Sigma_{-+}^{pp''} & -\delta^{pp''}\partial_u - \Sigma_{--}^{pp''} \end{pmatrix} \circ \begin{pmatrix} G_{++}^{p''p'} & G_{+-}^{p''p'} \\ G_{-+}^{p''p'} & G_{--}^{p''p'} \end{pmatrix} = \hat{I}.$$

$$\Sigma_{\eta\eta'}^{pp'} = -\eta\eta' \frac{\mathcal{J}^2}{q}(2G_{\eta\eta'}^{pp'})^{q-1}. \tag{23}$$

where $p, p', p'' \in \{a, b\}$ and $\bar{p} \neq p$. The large $q$ saddle-point equation can be derived after specifying $G_{0,\eta\eta'}^{pp'}$ with $\eta, \eta' = \pm$, which also takes the form of the Liouville equation with delta function sources. However, comparing to the single-sided calculation, there are two main difficulties: 1). unlike (15), the boundary condition in (22) breaks the time translation symmetry so there is no a simple way to find analytical solutions. 2). The $G_{0,\eta\eta'}^{pp'}$ takes different form in the short-time/long-time regime, as in the eternal traversable wormhole calculation [13]. In the following sections, we focus on the short-time limit and the long-time limit separately, where the Green's functions are almost translation invariant.

### 1. Short-time solution

For sufficiently short times $\hat{\mu}t \ll q$, the contribution due to a finite $\mu$ is perturbative. To leading order in $\hat{\mu}/\mathcal{J}$, the Green's function within each replica $a/b$ can be approximated by the equilibrium solution with $\hat{\mu} = 0$. Keeping to the $1/q$ order, the result reads [10]

$$G^D_{\eta\eta'}(u,u') = \langle \chi^{a/b}_\eta(u)\chi^{a/b}_{\eta'}(u') \rangle = \frac{G_{0,\eta\eta'}(\Delta u)}{\cosh^{2/q}(\mathcal{J}\Delta u)}. \quad (24)$$

For the inter-replica Green's functions, the result can be computed by solving (23) perturbatively in $\mu$. We find

$$\begin{aligned}G^{ba}_{\eta\eta'}(u,u') &= \langle \chi^b_\eta(u)\chi^a_{\eta'}(u') \rangle = \;_{\eta'}\!\!\underset{\mu}{\text{---}\bullet\text{---}}\!\!_{\eta} \\ &\approx \frac{\hat{\mu}}{q}\sum_\zeta \int_0^{2t} du''\, G^D_{\eta\zeta}(u,u'')G^D_{-\zeta\eta'}(u'',u').\end{aligned} \quad (25)$$

To $1/q$ order, we can replace $G^D_{\eta\eta'}$ by $G_{0,\eta\eta'}$. The result reads

$$G^{ba}_{\eta\eta'}(u,u') = \frac{\hat{\mu}}{2q}\begin{pmatrix} u+u'-2t & 2t-|u-u'| \\ 2t-|u-u'| & u+u'-2t \end{pmatrix}, \quad (26)$$

where $u,u' \in [0,2t]$. We find off-diagonal components are of the order $\hat{\mu}t/q$. It becomes large enough at sufficient long time $t \sim q$, where our perturbation results breakdown. By computing the on-shell action, it is straightforward to show that Eq. (26) results in $S \approx 2\hat{\mu}Nt/q$ for short time.

### 2. Long-time solution

At long times, the pairing between branches changes and the solution is highly replica off-diagonal. In the long-time limit $\hat{\mu}t \to \infty$, the Green's functions can be approximated by the "factorization of twist operators":

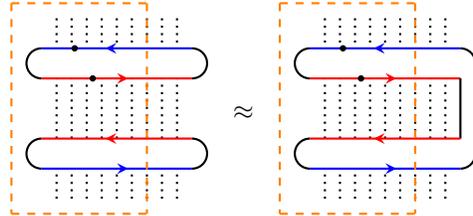

$$(27)$$

Here the black dots represent the insertion of Majorana operators and the orange dashed box indicates the region in which Green's functions are the same. Green's functions on the R.H.S. can be computed analytically. We first consider the Greens' functions where two fermion operators are inserted on branches $C_{\rm red} = (b,+) \cup (a,-)$. By contracting the redundant branches, we find

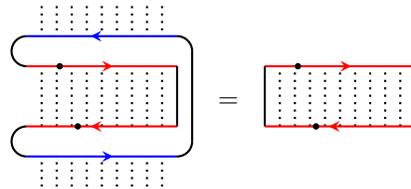

$$(28)$$

This is exactly the Keldysh contour (15) for the evolution of the density matrix. Consequently, the Green's function on $C_{\rm red}$ matches the equilibrium Green's functions (19) and (21) with finite $\hat{\mu}$:

$$\begin{aligned} G^{bb}_{++} &= -G^{aa}_{--} = \frac{1}{2}\mathrm{sgn}(\Delta u)\left[\frac{A}{\cosh(B+A\mathcal{J}|\Delta u|)}\right]^{2/q}, \\ G^{ba}_{+-} &= -G^{ab}_{-+} = \frac{1}{2}\left[\frac{A}{\cosh(B+A\mathcal{J}|\Delta u|)}\right]^{2/q}. \end{aligned} \quad (29)$$

Similar solutions hold on branches $C_{\text{blue}} = (b, -) \cup (a, +)$.

The remaining task is to compute the correlation between $C_{\text{blue}}$ and $C_{\text{red}}$. We take the two-point function illustrated in (27) as an example, while other components can be obtained straightforwardly. As in the previous calculations, we expand $G^{aa}_{-+} = \frac{1}{2}(1 + g^{aa}_{-+}/q + ...)$, the equation satisfied by $g^{aa}_{-+}$ is again of Liouville type,

$$\partial_u \partial_{u'} g^{aa}_{-+} = 2\mathcal{J}^2 e^{g^{aa}_{-+}}. \tag{30}$$

However, unlike (18), there is no source term induced by the environment. As we will see, this leads to a decay of $g^{aa}_{-+}(u, u')$ when $u$ and $u'$ are away from 0.

The general solution to the Liouville equation (30) can be written as

$$e^{g^{aa}_{-+}(u,u')} = -\frac{\mathcal{F}_1'(\mathcal{J}u)\mathcal{F}_2'(\mathcal{J}u')}{[\mathcal{F}_1(\mathcal{J}u) - \mathcal{F}_2(\mathcal{J}u')]^2}, \tag{31}$$

for arbitrary functions $(\mathcal{F}_1, \mathcal{F}_2)$. Using the previous boundary condition at $t = 0$, we have $(u > 0)$:

$$e^{g^{aa}_{-+}(0,u)} = e^{g^{aa}_{-+}(u,0)} = \frac{A^2}{\cosh^2(B + A\mathcal{J}u)}. \tag{32}$$

Combining (31) and (32), we obtain

$$\begin{aligned}
\mathcal{F}_1(x) &= \frac{1}{A \coth(Ax) + 2A \tanh B}, \\
\mathcal{F}_2(x) &= \frac{\coth(B + Ax)}{A}.
\end{aligned} \tag{33}$$

This gives $G^{aa}_{-+} = e^{g^{aa}_{-+}/q}/2$ with

$$e^{g^{aa}_{-+}} = \frac{A^2 \operatorname{csch}^2(A\mathcal{J}u)\operatorname{csch}^2(A\mathcal{J}u' + B)}{\{\coth(A\mathcal{J}u' + B)[\coth(A\mathcal{J}u) + 2\tanh B] - 1\}^2} \tag{34}$$

which determines the correlation between $C_{\text{blue}}$ and $C_{\text{red}}$. We observed the expected exponential decay for $\mathcal{J}u \gg 1$.

Furthermore, using the large-$q$ solution, we show that the long-time entropy $S = S^{(2)}_{\text{eff}}[\mathcal{G}, \Sigma] - 2S^{(1)}_{\text{eff}}[\mathcal{G}, \Sigma]$ is independent of $\mathcal{J}$ or $\hat{\mu}$ by computing its derivative with respect ot these variables $\partial_\mathcal{J} S = \partial_{\hat{\mu}} S = 0$. Since $S^{(1)}_{\text{eff}} = \operatorname{Tr}(\mathcal{P}) = Z_\beta = 2^{N/2}$ for $\beta = 0$, we focus on $\partial_\mathcal{J} S^{(2)}_{\text{eff}}$ and $\partial_{\hat{\mu}} S^{(2)}_{\text{eff}}$. Explicitly, we have

$$\partial_\mathcal{J} S^{(2)}_{\text{eff}}[\mathcal{G}, \Sigma] = -\frac{N\mathcal{J}}{2q^2} \int_\mathcal{C} du\, du'\ f(u)f(u')|2G(u,u')|^q = -\frac{N\mathcal{J}}{2q^2} \int_\mathcal{C} du\, du'\ f(u)f(u')e^{g(u,u')}. \tag{35}$$

Here we omit $a/b$ and $\pm$ labels for simplicity. Since Eq. (35) is local in time, (29) and (34) contribute separately. The contribution from (29) is zero. This is because that it matches the on-shell action of the single-sided contour (28). On the other hand, we know that the on-shell action of the single-sided contour is time independent due to the unitarity. Mathematically, this originates from the cancellation between contributions of $(G^{ab}_{-+}, G^{ba}_{+-})$ and $(G^{aa}_{--}, G^{bb}_{++})$, which takes different sign factors $f(u)f(u')$. Closer examination reveals similar cancellations exist for contributions from (34). As a result, we find $\partial_\mathcal{J} S^{(2)}_{\text{eff}} = 0$. We then compute

$$\partial_{\hat{\mu}} S^{(2)}_{\text{eff}}[\mathcal{G}, \Sigma] = -\frac{N}{2q} \int_\mathcal{C} du\, du' G(u,u')g(u,u') + \frac{mNt}{q}. \tag{36}$$

Following (10), $g(u, u')$ is a combination of Dirac delta functions. As a result, (36) only depends on $G^{ba}_{+-}(0) = -G^{ab}_{-+}(0) = 1/2$, which gives $\partial_{\hat{\mu}} S^{(2)}_{\text{eff}} = 0$. Finally, we expect $S = N \ln 2$ for $\hat{\mu}/\mathcal{J} \to \infty$ where the coupling to bath dominates. We thus conclude $S = N \ln 2$ in the long-time limit for arbitrary $\mathcal{J}/\hat{\mu}$.

---

* wanghanteng@sjtu.edu.cn
† cl91tp@gmail.com
‡ pengfeizhang.physics@gmail.com


\fi

\end{document}